\documentclass[apj]{emulateapj}
\usepackage{graphicx}
\usepackage{bm}
\usepackage{ulem}
\usepackage[dvips]{color}
\usepackage{xspace}
\usepackage{amsmath}

\renewcommand\thefootnote{\alpha{footnote}}

\shorttitle{Inflow and outflow rates at $z\sim1.4$
}
\shortauthors{Seko et al.}

\begin{document}
\title{Constraint on the inflow/outflow rates in star-forming galaxies at $z\sim1.4$ from molecular gas observations}

\author{
 Akifumi Seko\altaffilmark{1}, 
 Kouji Ohta\altaffilmark{1}, 
 Kiyoto Yabe\altaffilmark{2}, 
 Bunyo Hatsukade\altaffilmark{3}, 
 Masayuki Akiyama\altaffilmark{4}, 
 Naoyuki Tamura\altaffilmark{2}, 
 Fumihide Iwamuro\altaffilmark{1}, 
 and
 Gavin Dalton\altaffilmark{5,6}
}

\altaffiltext{1}{Department of Astronomy, Kyoto University, Kitashirakawa-Oiwake-Cho, Sakyo-ku, Kyoto, 606-8502, Japan}
\altaffiltext{2}{Kavli Institute for the Physics and Mathematics of the Universe, Todai Institutes for Advanced Study, 
the University of Tokyo, Kashiwa, 277-8583, Japan (Kavli IPMU, WPI)}
\altaffiltext{3}{National Astronomical Observatory of Japan, 2-21-1, Osawa, Mitaka, Tokyo, 181-8588, Japan}
\altaffiltext{4}{Astronomical Institute, Tohoku University, 6-3 Aramaki, Aoba-ku, Sendai, 980-8578, Japan}
\altaffiltext{5}{Department of Astrophysics, University of Oxford, Keble Road, Oxford OX1 3RH, UK}
\altaffiltext{6}{STFC RALSpace, Harwell, Oxfordshire OX11 0QX, UK}

\begin{abstract} 

We constrain the rate of gas inflow into and outflow from a main-sequence star-forming galaxy 
at $z\sim1.4$ by fitting a simple analytic model for the chemical evolution in a galaxy 
to the observational data of the stellar mass, metallicity, and molecular gas mass fraction. 
The molecular gas mass is derived from CO observations 
with a metallicity-dependent CO-to-H$_2$ conversion factor, 
and the gas metallicity is derived from the H$\alpha$ and [NII]$\lambda$~6584 emission line ratio. 
Using a stacking analysis of CO integrated intensity maps and the emission lines of H$\alpha$ and [NII], 
the relation between stellar mass, metallicity, and gas mass fraction is derived. 
We constrain the inflow and outflow rates with least-chi-square fitting 
of a simple analytic chemical evolution model to the observational data.
The best-fit inflow and outflow rates are $\sim$1.7 and $\sim$0.4 
in units of star-formation rate, respectively. 
The inflow rate is roughly comparable to the sum of the star-formation rate and outflow rate, 
which supports the equilibrium model for galaxy evolution; 
i.e., all inflow gas is consumed by star formation and outflow.

\end{abstract}


\section{Introduction}
Gas inflow and outflow play a very important role in galaxy evolution. 
There is indirect evidence for the existence of gas inflow. 
Firstly, the difference in the abundance distribution between observations and 
the closed-box model prediction \citep[e.g.,][]{Van62}, called the G-dwarf problem, 
which can be explained by the inflow of primordial gas \citep[e.g.,][]{Lars72}. 
Further evidence is that the timescale of gas depletion in star-forming galaxies 
at low redshift \citep[e.g.,][]{Wong02, Sain11} 
and high redshift \citep[e.g.,][]{Tacc10, Tacc13, Seko16}
is significantly shorter than that for building up their stellar masses \citep{Bouc10}, 
thus requiring gas inflow to sustain their star-formation activity. 
Gas outflow is found in local \citep[e.g.,][]{Sala13} 
and distant \citep[e.g.,][]{Wein09, Stei10, Genz11} star-forming galaxies.                                                                                                                                                                                                                                                                                                                                                                                                                                                                                                                                                                                                                                                                                                                                                                                                                                                                                                                                                                                                                                                                                                                                                                                                                                                                                       
\citet{Wein09} found that the outflow is ubiquitous at $z\sim1.4$ 
and the outflow rate is in the same order of magnitude as the star-formation rate (SFR) in galaxies.

Because inflow and outflow affect the gas mass (and its fraction) and gas-phase metallicity in a galaxy, 
efforts have been made to constrain the inflow and outflow rates 
to reproduce the observational relations such as the stellar mass-metallicity relation 
by using cosmological simulations \citep[e.g.,][]{Kere05, Finl08} 
and analytic models \citep[e.g.,][]{Bouc10, Lill13}. 
Some of the studies showed that galaxies evolve while maintaining the balance between the amounts of inflow gas, 
star formation, and outflow: inflow $=$ star formation $+$ outflow. 
Such scenario is called the ``equilibrium model'' for galaxy evolution. 
By using near-infrared spectroscopy of star-forming galaxies at $z\sim2$ \citep{Erb06a}, 
\citet{Erb08} derived gas-phase metallicities from emission lines 
and gas mass fractions from extinction corrected H$\alpha$ luminosities 
by assuming the Kennicutt-Schmidt law. 
Then, she derived the inflow and outflow rates at $z\sim2$ by fitting a simple analytic model 
for the chemical evolution in a galaxy to these quantities. 
Her result supports the equilibrium model. 
\citet{Yabe15} tried to constrain the cosmic evolution of inflow and outflow rates 
by using the chemical evolution model and their near-infrared spectroscopic data 
at $z\sim1.4$ \citep{Yabe14} and observational data at $z\sim0$ and 2 
in literature \citep[][respectively]{Peep11, Erb06a}. 
They also derived gas-phase metallicities from emission lines and 
gas mass fractions at $z\sim1.4$ and 2 from extinction corrected H$\alpha$ luminosities 
by assuming the Kennicutt-Schmidt law. 
They found the inflow, outflow, and star-formation rates decreased while satisfying 
the equilibrium condition at all the redshifts. 
However, no studies constrain the inflow and outflow rates at $z>1$ with chemical evolution models 
using molecular gas observations for the gas mass fraction.

The advent of high-sensitivity radio telescopes recently enables us to detect CO emissions 
from ``normal'' star-forming galaxies at $z>1$. 
The stellar masses of most star-forming galaxies are well correlated with SFR at each redshift, 
and galaxies located on this correlation are referred to as ``main-sequence galaxies'' 
\citep[e.g.,][]{Dadd07, Noes07}. 
CO observations of main-sequence galaxies at $z=1$-1.5 were conducted 
by covering a wide range of stellar mass \citep[e.g.,][]{Tacc13, Seko16} 
and metallicity \citep[e.g.,][]{Seko16}. 
The relations between stellar mass, metallicity, and (molecular) gas mass are gradually being revealed 
by these studies. 
In this paper, we constrain the inflow and outflow rates 
from the results of molecular gas observations at $z\sim1.4$ 
by using the simple chemical evolution model. 
We describe our sample and observational data in section~\ref{sec: sample}. 
In section~\ref{sec: CEM} and \ref{sec: results}, 
the chemical evolution model used and fitting results are described, respectively. 
We present discussions in section~\ref{sec: discussions}, and a summary in section~\ref{sec: summary}. 
Throughout this paper, we adopt the standard $\Lambda$-CDM cosmology 
with $H_0=70~\mathrm{km~s^{-1}~Mpc^{-1}}$, $\Omega_M=0.3$, and $\Omega_\lambda=0.7$.

\begin{figure*}[t]
\includegraphics[width = 17.5cm]{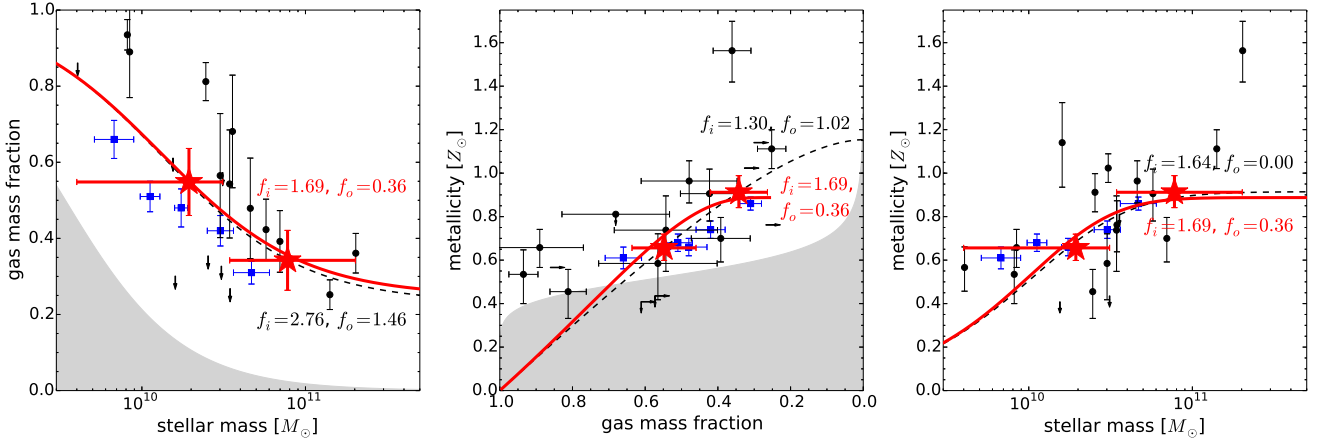}
\caption{Diagrams of gas mass fraction versus stellar mass (left), 
metallicity versus gas mass fraction (middle), 
and metallicity versus stellar mass (right). 
Red stars represent the results of the stacking analysis for the subsamples with smaller/larger stellar mass. 
Black circles and arrows refer to the results for individual galaxies from \citet{Seko16}. 
Blue squares represent the results of \citet{Yabe15}, 
who derived the gas mass fractions by using the Kennicutt-Schmidt law. 
The red solid line shows the best-fit model derived from the joint fitting. 
The black dashed line shows the best-fit model derived from fitting in each panel. 
The gray shaded region represents our CO observational limit (see the text for details). }
\label{fig: fitting}
\end{figure*}

\section{Sample and observational data}\label{sec: sample}
\subsection{NIR spectroscopic sample at $z\sim1.4$}
The sample used in this paper comes originally from $K_{s}$-band selected galaxies 
in the Subaru/XMM-Newton Deep Survey (SXDS) field. 
\citet{Yabe12} derived photometric redshift and stellar mass 
by fitting population synthesis models to the spectral energy distributions (SEDs) 
constructed with multi-wavelength data, from ultra-violet (UV) to near-infrared. 
SFRs were derived from rest-frame UV luminosity densities, 
corrected for the dust extinction estimated from the rest-frame UV spectral slope. 
In both stellar mass and SFR, the Salpeter initial mass function \citep[IMF;][]{Salp55} 
with a mass range of $0.1-100~M_\odot$ was adopted. 

\citet{Yabe12} selected star-forming galaxies at $z_\mathrm{phot}\sim1.4$ with $K_{s}<23.9~\mathrm{mag}$
(AB magnitude), stellar mass $\geq 10^{9.5}~M_\odot$, and expected H$\alpha$ flux
$\geq10^{-16}~\mathrm{erg~s^{-1}~cm^{-2}}$. 
Most of the selected galaxies are located on the main sequence at this redshift. 
They obtained near-infrared spectroscopy of the selected galaxies with the Fiber Multi-Object 
Spectrograph (FMOS) on the Subaru telescope and detected H$\alpha$ emission 
from 71 out of 317 observed galaxies. 
The metallicity was derived using the H$\alpha$ and [NII]$\lambda~6584$ emission line ratio 
\citep[N2 method;][]{PP04}.

\subsection{CO observations with ALMA}
\citet{Seko16} selected 20 galaxies from the H$\alpha$ sample mentioned above 
to trace the distributions in diagrams of SFR versus stellar mass 
and metallicity versus stellar mass with good uniformity. 
The ranges of stellar mass and metallicity 
in the selected galaxies are $4\times10^{9}$--$4\times10^{11}~M_\odot$ and 0.3--1.5~$Z_\odot$ 
\citep[the solar oxygen abundance $Z_\odot = 12+\log(\mathrm{O/H}) = 8.66$;][]{Aspl04}, respectively. 
They conducted $^{12}$CO($J=5$-4) observations toward 20 star-forming galaxies 
(two galaxies are not located on the main sequence) with 
Atacama Large Millimeter/submillimeter Array (ALMA) during the ALMA cycle0 session 
(ID=2011.0.00648.S, PI=K. Ohta). 
CO emission was detected in 11 galaxies. 

\subsection{Stacking analysis}
We separated the sample galaxies into two subsamples with smaller ($<10^{10.5}~M_\odot$) 
and larger ($>10^{10.5}~M_\odot$) stellar mass, except for the two non-main-sequence galaxies. 
The average stellar mass is $1.9\times10^{10}~M_\odot$ and $7.8\times10^{10}~M_\odot$ 
for the subsamples with smaller and larger stellar mass, respectively. 
Then, we carried out an image-based stacking analysis to derive the average value
(details are described by \citet{Seko16}), 
since the CO emissions from about half of our sample galaxies are not detected. 

The CO($J=5$-4) line luminosity ($L_\mathrm{CO(5-4)}^{'}$) is given as 
\begin{equation}
L_\mathrm{CO(5-4)}^{'} = 3.25\times10^{7} S_\mathrm{CO(5-4)} \Delta v \nu_\mathrm{rest(5-4)}^{-2} D_L^2 (1+z)^{-1}, 
\end{equation}
where $L_\mathrm{CO(5-4)}^{'}$ is measured in $\mathrm{K~km~s^{-1}~pc^2}$, 
$S_\mathrm{CO(5-4)} \Delta v$ is the observed CO(5-4) integrated flux density 
in $\mathrm{Jy~km~s^{-1}}$, 
$\nu_\mathrm{rest(5-4)}$ is the rest-frame frequency of the CO(5-4) emission line in $\mathrm{GHz}$, 
and $D_L$ is the luminosity distance in $\mathrm{Mpc}$. 
The resulting CO(5-4) line luminosities for the subsamples with smaller and larger stellar mass are 
$(9.3\pm3.0)\times10^{8}~\mathrm{K~km~s^{-1}~pc^2}$ and 
$(2.3\pm0.5)\times10^{9}~\mathrm{K~km~s^{-1}~pc^2}$, respectively.  
The errors are derived from a random resampling of stacked galaxies. 
To derive the average metallicities of the subsamples, 
we also conducted a stacking analysis of the near-infrared spectra obtained with FMOS
(details are described by \citet{Yabe12, Yabe14}). 
The resulting metallicities for the subsamples with smaller and larger stellar mass 
are $0.66\pm0.06~Z_\odot$ and $0.91\pm0.07~Z_\odot$, respectively. 

\subsection{Molecular gas mass and its fraction}
To derive the molecular gas mass, 
we adopted a luminosity ratio ($L_\mathrm{CO(5-4)}^{'}/L_\mathrm{CO(1-0)}^{'}$) of 0.23, 
which is the average value of three sBzK galaxies at $z\sim1.5$ \citep{Dadd15}.
We adopted the metallicity-dependent 
CO-to-H$_2$ conversion factor given by \citet{Genz12}. 
Because \citet{Genz12} used the metallicity calibration of \citet{Deni02}, 
we converted the metallicity using an empirical relation between the two metallicity calibrations 
obtained by \citet{Kwel08}. 
The resultant molecular gas masses for the subsamples with smaller and larger stellar mass are 
$(2.4\pm0.7)\times10^{10}~M_\odot$ and $(4.0\pm1.0)\times10^{10}~M_\odot$, respectively, and 
the gas mass fractions ($M_\mathrm{mol}/(M_\mathrm{mol} + M_\ast)$, including the helium mass 
for $M_\mathrm{mol}$) are $0.55\pm0.09$ and $0.34\pm0.08$, respectively. 
The stellar masses, gas mass fractions, and metallicities of the subsamples are plotted 
with red stars in Figure~\ref{fig: fitting} together with individual galaxies (black symbols); 
the left, middle, and right panel are diagrams of 
gas mass fraction versus stellar mass ($\mu$ vs $M_\ast$), 
metallicity versus gas mass fraction ($Z$ vs $\mu$), 
and metallicity versus stellar mass ($Z$ vs $M_\ast$), respectively. 
In figure~\ref{fig: fitting}, the results for $\sim$340 star-forming galaxies at $z\sim1.4$ 
by \citet{Yabe15} are also plotted (blue squares) and are consistent with the results obtained in this study. 
They stacked the near-infrared spectroscopic data 
for the subsamples separated into five stellar mass bins. 
They derived metallicities with the N2 method 
and gas mass fractions from the H$\alpha$ luminosities and sizes of galaxies 
by applying the Kennicutt-Schmidt law. 

The observational limits of the gas mass fraction are shown 
in the left and middle panels of Figure~\ref{fig: fitting} with gray shades. 
The noise level of the CO stacking analysis for both subsamples ($\sim0.25~\mathrm{mJy}$) 
gives an upper limit of the CO(1-0) luminosity of $\sim4.3\times10^{8}~\mathrm{K~km~s^{-1}~pc^2}$ 
by assuming a velocity width of $200~\mathrm{km~s^{-1}}$, 
which is the average full width at half maximum of the detected CO emission lines from \citet{Seko16}, 
and the same CO luminosity ratio above. 
This limit leads to an upper gas mass fraction at a fixed stellar mass 
(left panel of Figure~\ref{fig: fitting}) and at a fixed metallicity (middle panel of Figure~\ref{fig: fitting}), 
if we assume the stellar mass-metallicity relation at $z\sim1.4$ \citep{Yabe14} 
and the metallicity-dependent CO-to-H$_2$ conversion factor \citep{Genz12}. 
In the middle panel of Figure~\ref{fig: fitting}, 
the upper limit of the gas mass fraction is very high at lower metallicity. 
This is because the stellar mass-metallicity relation we adopted 
implies a very small stellar mass for the lower metallicity galaxies, resulting into a very large gas mass fraction. 
A few of our galaxies are located in the gray shaded region in the middle panel of Figure~\ref{fig: fitting}. 
This is because they show large deviations from the assumed stellar mass-metallicity relation.

\section{Chemical evolution model}\label{sec: CEM}
We compare the observational data with an analytic model for chemical evolution 
in a galaxy considering the gas inflow and outflow. 
Based on the description by \citet{Matt01}, 
the equation for the evolution of gas is given as 
\begin{equation}
\frac{dM_\mathrm{gas}}{dt} = -(1-R)\psi(t) + A(t) - W(t), 
\end{equation}
where $R$ is the total mass fraction which is restored to the interstellar medium by a stellar generation, 
$\psi(t)$ is SFR, $A(t)$ is the inflow (accretion) rate, and $W(t)$ is the outflow (wind) rate. 
The evolution of metal mass in gas phase is given as 
\begin{equation}
\frac{d(ZM_\mathrm{gas})}{dt} = -(1-R)Z(t)\psi(t) + y_{Z}(1-R)\psi(t) + Z_{A}A(t) - ZW(t), 
\end{equation}
where $Z$ is the gas phase metallicity, $y_{Z}$ is the yield, and $Z_{A}$ is the metallicity of inflow gas. 
These equations are given by adopting the instantaneous recycling approximation. 
We adopted $y_{Z}=1.5~Z_\odot$ in this work, 
which is the value used by \citet{Erb08} and \citet{Yabe15}.
The inflow rate and outflow rate are assumed to be proportional to the SFR such as 
$A(t)=f_{i}(1-R)\psi(t)$ and $W(t)=f_{o}(1-R)\psi(t)$, respectively, 
and the inflow gas is assumed to be primordial ($Z_{A}=0$). 
The assumption of an outflow rate proportional to the SFR would be appropriate 
if the galactic winds are mainly driven by supernovae explosions.  
The assumption of an inflow rate proportional to SFR would be reasonable 
if the amount of inflow gas is closely related to the mass of gas available to form stars. 
These equations analytically lead to 
\begin{equation}
Z = \frac{y_{Z}}{f_{i}}(1 - [(f_{i} - f_{o}) - (f_{i} - f_{o} - 1)\mu^{-1}]^{\frac{f_{i}}{f_{i} - f_{o} - 1}}), 
\label{eq: mu vs Z}
\end{equation}
and $\mu$ is written as, 
\begin{equation}
\mu = \frac{M_\mathrm{gas}^{0} + (f_{i} - f_{o} - 1) M_\ast}{M_\mathrm{gas}^{0} + (f_{i} - f_{o}) M_\ast}, 
\label{eq: mu vs Mstar}
\end{equation}
where $M_\mathrm{gas}^{0}$ is the initial mass of primordial gas in a galaxy. 

\section{Results}\label{sec: results}
\begin{figure}[t]
\hspace{-0.7cm}
\includegraphics[width = 10cm]{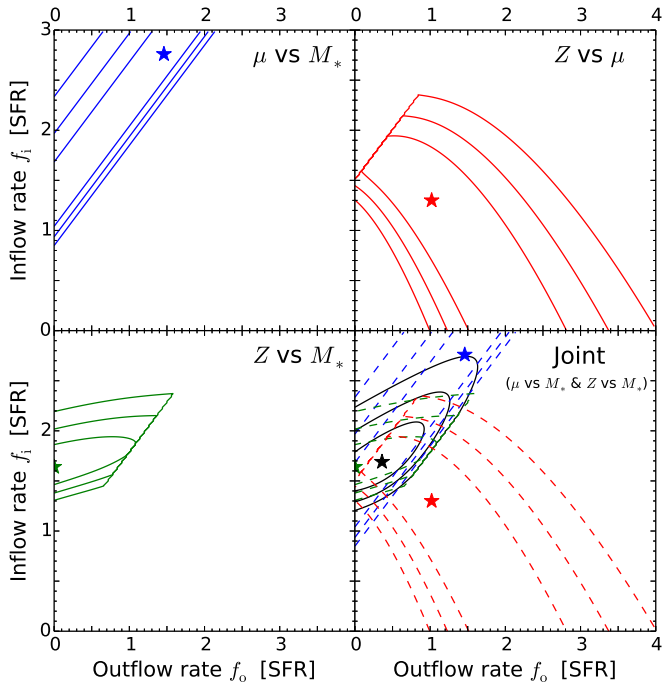}
\caption{$\chi^{2}$ contour maps of the fits for the diagrams of gas mass fraction 
versus stellar mass (top left), metallicity versus gas mass fraction (top right), 
metallicity versus stellar mass (bottom left), and the joint fit with the diagrams of 
gas mass fraction versus stellar mass and metallicity versus stellar mass (bottom right). 
Stars show the best-fit value in each panel. 
Contours represent 1$\sigma$, 2$\sigma$, and 3$\sigma$.}
\label{fig: chi2}
\end{figure}

To constrain the inflow ($f_{i}$) and outflow ($f_{o}$) rates, 
the analytic solutions are fitted to the observational data in each diagram of Figure~\ref{fig: fitting}. 
As shown in these equations, $f_{i}$, $f_{o}$, and $M_\mathrm{gas}^{0}$ are free parameters. 
From the least chi-square fitting in the diagram of $\mu$ vs $M_\ast$, 
we obtain a value of $M_\mathrm{gas}^{0} = 10^{10.24}~M_\odot$. 
The ranges of the parameters examined are $\log(M_\mathrm{gas}^{0}/M_\odot)=9.0$-11.0, 
$f_{i}=0.0$-3.0, and $f_{o}=0.0$-4.0 with 50 grids for each parameter. 
Then, we fixed the initial gas mass and fit in each diagram of Figure~\ref{fig: fitting} 
in the parameter ranges of $f_{i}=0.0$-3.0 and $f_{o}=0.0$-4.0 with 200 grids.

In Figure~\ref{fig: chi2}, the $\chi^{2}$ contour maps of the inflow and outflow rates are shown; 
the top left, top right, and bottom left panel show the fits for 
$\mu$ vs $M_\ast$, $Z$ vs $\mu$, and $Z$ vs $M_\ast$, respectively. 
The best-fit combinations of inflow and outflow rates ($f_i$, $f_o$) are
($2.76^{+0.23}_{-1.71}$, $1.46^{+0.48}_{-1.46}$), 
($1.30^{+0.65}_{-1.30}$, $1.02^{+1.80}_{-0.94}$), 
and ($1.64^{+0.33}_{-0.18}$, $0.00^{+1.14}_{-0.00}$) 
for the $\mu$ vs $M_\ast$, $Z$ vs $\mu$, and $Z$ vs $M_\ast$ diagram, respectively. 
The best-fit model is plotted with a black dashed line in each panel in Figure~\ref{fig: fitting} and 
the best-fit combination of inflow and outflow rates is represented 
with a colored star in each panel in Figure~\ref{fig: chi2}. 
Although the degeneracy between the inflow and outflow rates is severe in the fitting of each diagram, 
since the directions of the degeneracy in each diagram are different, 
the inflow and outflow rates can be constrained with a joint fit. 
Since the three relations shown in Figure~\ref{fig: fitting} are not independent, 
in this paper, we adopt the inflow and outflow rates from the joint fit from 
$\mu$ vs $M_\ast$ and $Z$ vs $M_\ast$, as done by \citet{Yabe15}. 
The results of joint fitting  of $\mu$ vs $M_\ast$ and $Z$ vs $\mu$, 
and $Z$ vs $\mu$ and $Z$ vs $M_\ast$ are almost the same 
as that for $\mu$ vs $M_\ast$ and $Z$ vs $M_\ast$. 
We explore the combination of inflow and outflow rates which minimizes the joint chi-square. 
The best-fit values are ($f_i$, $f_o$)=($1.69^{+0.44}_{-0.30}$, $0.36^{+0.60}_{-0.36}$) 
which are shown with a black star in the bottom right panel of Figure~\ref{fig: chi2}. 
The fit result is shown with red solid line in Figure~\ref{fig: fitting}.

\section{Discussions}\label{sec: discussions}
As mentioned above, \citet{Yabe15} constrain the inflow and outflow rates at the same redshift. 
Our result is consistent with that of \citet{Yabe15} 
(($f_{i}$, $f_{o}$$) = ($$1.84^{+0.14}_{-0.14}$, $0.60^{+0.18}_{-0.22}$)) within the errors. 
Although the uncertainty of the inflow and outflow rate is large, 
the best-fit model shows that the inflow rate is roughly equal to the sum of 
the outflow rate and effective star-formation rate ($(1-R)\psi$), 
which supports the equilibrium model for galaxy evolution.  
It should be noted that the inflow and outflow rates are $1.47^{+0.45}_{-0.23}$ and 
$0.04^{+0.76}_{-0.04}$, respectively, if we adopt the Chabrier IMF \citep{Chab03}. 
The stellar mass with the Chabrier IMF is converted from that with the Salpeter IMF 
by dividing by 1.7 \citep[e.g.,][]{Spea14}. 
While the outflow rate is smaller than that derived with the Salpeter IMF, 
the equilibrium model for galaxy evolution is still supported.  

We also examine the inflow and outflow rates by using the subsamples 
with lower ($<8.5$) and higher ($>8.5$) metallicity (details are described by \citet{Seko16}). 
The average stellar masses are $1.7\times10^{10}~M_\odot$ and $6.6\times10^{10}~M_\odot$ 
for the subsamples with lower and higher metallicity, respectively. 
The stacked molecular gas mass fractions for the subsamples with lower and higher metallicity are 
$0.65\pm0.07$ and $0.32\pm0.09$, respectively. 
The stacked metallicities for the subsamples with lower and higher metallicity are 
$0.55\pm0.05~Z_\odot$ and $0.95\pm0.07~Z_\odot$, respectively. 
The joint fit gives an inflow rate of $1.50^{+0.41}_{-0.34}$ 
and outflow rate of $0.54^{+0.58}_{-0.54}$, which are consistent with the result 
derived by using the subsamples based on stellar mass, within the errors. 
The result also supports the equilibrium model for galaxy evolution. 

According to the study of CO luminosity ratios for main-sequence galaxies at $z\sim1.5$ by \citet{Dadd15}, 
the uncertainty of the luminosity ratio ($L_\mathrm{CO(5-4)}^{'}/L_\mathrm{CO(1-0)}^{'}$) is 
about a factor of 2. 
If we adopt the luminosity ratio of 0.12, 
the inflow and outflow rates are $1.40^{+0.46}_{-0.21}$ and $0.00^{+0.74}_{-0.00}$ 
($M_\mathrm{gas}^{0} = 10^{10.56}~M_\odot$), respectively. 
If we adopt the luminosity ratio of 0.46, 
the inflow and outflow rates are $1.92^{+0.40}_{-0.30}$ and $0.76^{+0.50}_{-0.56}$ 
($M_\mathrm{gas}^{0} = 10^{9.96}~M_\odot$), respectively. 
As the luminosity ratio is larger (i.e., molecular gas mass fraction is smaller), 
the inflow and outflow rates become large but the initial gas mass becomes small. 

If we adopt the Galactic CO-to-H$_2$ conversion factor 
($4.36~M_\odot~(\mathrm{km~s^{-1}~pc^2})^{-1}$) to derive the molecular gas mass, 
the inflow and outflow rates are $1.86^{+0.42}_{-0.30}$ and $0.40^{+0.60}_{-0.40}$ 
for the subsamples with smaller and larger stellar mass, 
and $1.88^{+0.46}_{-0.34}$ and $0.42^{+0.72}_{-0.42}$ 
for the subsamples with lower and higher metallicity. 
These best-fit values are consistent with those derived 
with the metallicity-dependent CO-to-H$_2$ conversion factor. 
However, the $\chi^2$ of the joint fit is worse, and the best-fit models do not reproduce 
the observational data well in the diagram of $Z$ vs $\mu$ which is not used for the joint fit. 

Although the gas mass in the chemical evolution model includes the HI mass, 
the gas mass fraction of observational data used in this paper does not include it. 
According to a semi-empirical model by \citet{Popp15}, the
HI mass in a galaxy at $z\sim1.4$ with a halo mass of $10^{12-14}~M_\odot$ is 
half or comparable to the H$_2$ mass. 
We examine the inflow and outflow rate assuming two cases that 
the HI mass in a galaxy is half of the H$_2$ mass and the same amount of the H$_2$ mass 
in a galaxy for the subsample with smaller/larger stellar mass. 
In these cases, the gas mass fraction is defined as 
$\mu = (M(\mathrm{H_2}) + M(\mathrm{HI}))/(M(\mathrm{H_2}) + M(\mathrm{HI}) + M_\ast)$. 
The best-fit inflow and outflow rates for the case of $M(\mathrm{HI})=0.5M(\mathrm{H_2})$ and 
the case of $M(\mathrm{HI})=M(\mathrm{H_2})$ are 
$1.51^{+0.44}_{-0.25}$ and $0.14^{+0.68}_{-0.14}$ 
($M_\mathrm{gas}^{0} = 10^{10.44}~M_\odot$) and 
$1.40^{+0.46}_{-0.21}$ and $0.00^{+0.74}_{-0.00}$ 
($M_\mathrm{gas}^{0} = 10^{10.56}~M_\odot$), respectively. 
The inflow and outflow rates are smaller and the initial gas mass is larger 
than those without HI mass.
If we assume a larger HI mass than these (i.e., $M(\mathrm{HI}) \geq 2M(\mathrm{H_2})$), 
the large amount of the initial gas mass causes a lower metallicity and 
the best-fit models in these cases are unable to reproduce the observed stacked values of metallicity. 

While the value of outflow rate is consistent with zero within the error, 
the small number of stacked data points leads to the large uncertainty. 
\citet{Heck02} showed galactic-scale superwinds exist in galaxies 
with a global SFR surface density exceeding 0.1~$M_\odot~\mathrm{yr^{-1}~kpc^{-2}}$ 
from observations of local starburst galaxies. 
Because the SFR surface density in all of our sample, whose size is derived from {\it B}-band image, 
exceeds the threshold, the outflow rates of our sample may not be zero.

\section{Summary}\label{sec: summary}
We constrain the inflow and outflow rates in star-forming galaxies at $z\sim1.4$ 
by using the analytic model for the chemical evolution in a galaxy. 
The gas mass fractions of our sample are derived 
from CO(5-4) observations with ALMA by adopting the metallicity-dependent 
CO-to-H$_2$ conversion factor. 
The joint least $\chi^2$ fit of the analytic model to the result of stacking analysis 
for the subsample with smaller/larger stellar mass 
shows the inflow and outflow rates in the unit of SFR 
are $1.69^{+0.44}_{-0.30}$ and $0.36^{+0.60}_{-0.36}$, respectively. 
The result is consistent with that from the previous study 
in which the gas mass was derived from extinction corrected H$\alpha$ luminosity and galaxy size 
by assuming the Kennicutt-Schmidt law. 
The inflow rate is roughly comparable to the sum of outflow rate and effective SFR, 
which supports the equilibrium model for galaxy evolution. 
The result is also consistent with that derived from the stacking analysis 
for subsamples with lower/higher metallicity. 

If we include the HI mass which is proportional to molecular gas mass, 
then lower inflow and outflow rates and larger initial gas masses are needed 
to explain the data. 
However, if we assume the amount of HI mass is larger than H$_2$ mass, 
the chemical evolution model used in this paper produces much lower metallicity, 
and do not reproduce the stacked metallicity. 

The uncertainty of inflow and outflow rates is large 
due to the small number of stacked data points. 
To obtain more reliable constraints of inflow and outflow rates,  
we need to increase the number of CO observations toward main-sequence galaxies 
with known metallicity covering wide ranges of stellar mass and metallicity.

\acknowledgments
We would like to thank the referee for useful comments and suggestions. 
A.S. is supported by Research Fellowship for Young Scientists 
from the Japan Society of the Promotion of Science (JSPS). 
K.O. is supported by Grant-in-Aid for Scientific Research (C) (16K05294) from JSPS. 
K.Y. is supported by Grant-in-Aid for Young Scientists (B) (JP16K17659) from JSPS.
Kavli IPMU is supported by World Premier International Research 
Center Initiative (WPI), MEXT, Japan.


\begin{thebibliography}{}

\bibitem[Asplund et al.(2004)]{Aspl04} Asplund, M., Grevesse, N., Sauval, A.~J., Allende Prieto, C., \& Kiselman, D.\ 2004, \aap, 417, 751 
\bibitem[Bouch{\'e} et al.(2010)]{Bouc10} Bouch{\'e}, N., Dekel, A., Genzel, R., et al.\ 2010, \apj, 718, 1001 
\bibitem[Chabrier(2003)]{Chab03} Chabrier, G.\ 2003, \pasp, 115, 763 
\bibitem[Daddi et al.(2015)]{Dadd15} Daddi, E., Dannerbauer, H., Liu, D., et al.\ 2015, \aap, 577, A46 
\bibitem[Daddi et al.(2007)]{Dadd07} Daddi, E., Dickinson, M., Morrison, G., et al.\ 2007, \apj, 670, 156 
\bibitem[Denicol{\'o} et al.(2002)]{Deni02} Denicol{\'o}, G., Terlevich, R., \& Terlevich, E.\ 2002, \mnras, 330, 69 
\bibitem[Erb(2008)]{Erb08} Erb, D.~K.\ 2008, \apj, 674, 151-156 
\bibitem[Erb et al.(2006)]{Erb06a} Erb, D.~K., Shapley, A.~E., Pettini, M., et al.\ 2006, \apj, 644, 813 
\bibitem[Finlator \& Dav{\'e}(2008)]{Finl08} Finlator, K., \& Dav{\'e}, R.\ 2008, \mnras, 385, 2181 
\bibitem[Genzel et al.(2011)]{Genz11} Genzel, R., Newman, S., Jones, T., et al.\ 2011, \apj, 733, 101 
\bibitem[Genzel et al.(2012)]{Genz12} Genzel, R., Tacconi, L.~J., Combes, F., et al.\ 2012, \apj, 746, 69 
\bibitem[Heckman(2002)]{Heck02} Heckman, T.~M.\ 2002, in ASP Conf. Ser. 254, 
Extragalactic Gas at Low Redshift, ed. J. S. Mulchaey \& J. Stocke (San Francisco, CA: ASP), 292 
\bibitem[Kere{\v s} et al.(2005)]{Kere05} Kere{\v s}, D., Katz, N., Weinberg, D.~H., \& Dav{\'e}, R.\ 2005, \mnras, 363, 2 
\bibitem[Kewley \& Ellison(2008)]{Kwel08} Kewley, L.~J., \& Ellison, S.~L.\ 2008, \apj, 681, 1183-1204 
\bibitem[Larson(1972)]{Lars72} Larson, R.~B.\ 1972, \nat, 236, 21 
\bibitem[Lilly et al.(2013)]{Lill13} Lilly, S.~J., Carollo, C.~M., Pipino, A., Renzini, A., \& Peng, Y.\ 2013, \apj, 772, 119 
\bibitem[Matteucci(2001)]{Matt01} Matteucci, F.\ 2001, Astrophysics and Space Science Library, 253,  
\bibitem[Noeske et al.(2007)]{Noes07} Noeske, K.~G., Weiner, B.~J., Faber, S.~M., et al.\ 2007, \apjl, 660, L43 
\bibitem[Peeples \& Shankar(2011)]{Peep11} Peeples, M.~S., \& Shankar, F.\ 2011, \mnras, 417, 2962 
\bibitem[Pettini \& Pagel(2004)]{PP04} Pettini, M., \& Pagel, B.~E.~J.\ 2004, \mnras, 348, L59 
\bibitem[Popping et al.(2015)]{Popp15} Popping, G., Behroozi, P.~S., \& Peeples, M.~S.\ 2015, \mnras, 449, 477 
\bibitem[Saintonge et al.(2011)]{Sain11} Saintonge, A., Kauffmann, G., Wang, J., et al.\ 2011, \mnras, 415, 61 
\bibitem[Salak et al.(2013)]{Sala13} Salak, D., Nakai, N., Miyamoto, Y., Yamauchi, A., \& Tsuru, T.~G.\ 2013, \pasj, 65,  
\bibitem[Salpeter(1955)]{Salp55} Salpeter, E.~E.\ 1955, \apj, 121, 161 
\bibitem[Seko et al.(2016)]{Seko16} Seko, A., Ohta, K., Yabe, K. et al.\ 2016, \apj, 819, 82 
\bibitem[Speagle et al.(2014)]{Spea14} Speagle, J.~S., Steinhardt, C.~L., Capak, P.~L., \& Silverman, J.~D.\ 2014, \apjs, 214, 15 
\bibitem[Steidel et al.(2010)]{Stei10} Steidel, C.~C., Erb, D.~K., Shapley, A.~E., et al.\ 2010, \apj, 717, 289 
\bibitem[Tacconi et al.(2010)]{Tacc10} Tacconi, L.~J., Genzel, R., Neri, R., et al.\ 2010, \nat, 463, 781 
\bibitem[Tacconi et al.(2013)]{Tacc13} Tacconi, L.~J., Neri, R., Genzel, R., et al.\ 2013, \apj, 768, 74 
\bibitem[van den Bergh(1962)]{Van62} van den Bergh, S.\ 1962, \aj, 67, 486 
\bibitem[Weiner et al.(2009)]{Wein09} Weiner, B.~J., Coil, A.~L., Prochaska, J.~X., et al.\ 2009, \apj, 692, 187 
\bibitem[Wong \& Blitz(2002)]{Wong02} Wong, T., \& Blitz, L.\ 2002, \apj, 569, 157 
\bibitem[Yabe et al.(2015)]{Yabe15} Yabe, K., Ohta, K., Akiyama, M., et al.\ 2015, \apj, 798, 45 
\bibitem[Yabe et al.(2012)]{Yabe12} Yabe, K., Ohta, K., Iwamuro, F., et al.\ 2012, \pasj, 64, 60 
\bibitem[Yabe et al.(2014)]{Yabe14} Yabe, K., Ohta, K., Iwamuro, F., et al.\ 2014, \mnras, 437, 3647 



\end{thebibliography}
\end{document}